\documentclass[letterpaper,  %a4paper
               keeplastbox,
              ]{jacow}
\makeatletter%
	\ifboolexpr{bool{xetex}}
	 {\renewcommand{\Gin@extensions}{.pdf,%
	                    .png,.jpg,.bmp,.pict,.tif,.psd,.mac,.sga,.tga,.gif,%
	                    .eps,.ps,%
	                    }}{}
\makeatother

\ifboolexpr{bool{xetex} or bool{luatex}} % test for XeTeX/LuaTeX
 {}                                      % input encoding is utf8 by default
 {\usepackage[utf8]{inputenc}}           % switch to utf8

\usepackage[USenglish]{babel}			 
\usepackage{graphicx}
\usepackage[final]{pdfpages}
\usepackage{multirow}
\usepackage{ragged2e}
\usepackage{dblfloatfix}
\usepackage{blindtext}
\usepackage{subfigure}  
\usepackage{lipsum}
\let\OLDthebibliography\thebibliography
\renewcommand\thebibliography[1]{
  \OLDthebibliography{#1}
  \setlength{\parskip}{0pt}
  \setlength{\itemsep}{0pt plus 0.3ex}
  }
\ifboolexpr{bool{jacowbiblatex}}%
 {%
  \addbibresource{jacow-test.bib}
  \addbibresource{biblatex-examples.bib}
 }{}
\begin{document}

\title{A Wedge Test in MICE\thanks{MICE has been made possible by grants from DOE, NSF (U.S.A), the INFN (Italy), the STFC (U.K.), the European Community under the European Commission Framework Programme 7, the Japan Society for the Promotion of Science, the Swiss National Science Foundation, and the Fermi Research Alliance, LLC operating Fermilab under contract No. DE-AC02-07CH11359 with the U.S. Department of Energy.}}

\author{Tanaz Angelina Mohayai\thanks{tmohayai@hawk.iit.edu}, Pavel Snopok, Illinois Institute of Technology, Chicago IL, USA \\  
	David Neuffer, Fermilab, Batavia IL, USA \\
	Don Summers, University of Mississippi, Oxford MS, USA \\
		on behalf of the MICE Collaboration
		}
	
\maketitle
\begin{abstract}
Emittance exchange mediated by wedge absorbers can be used for longitudinal ionization cooling and for final transverse emittance minimization for a muon collider. A wedge absorber within the Muon Ionization Cooling Experiment (MICE) could serve as a demonstration of the type of emittance exchange needed for six-dimensional (6D) cooling, including the configurations needed for muon colliders. Parameters for this test have been explored in simulation and applied to experimental configurations using a wedge-shaped absorber. A polyethylene wedge absorber has been fabricated and placed in MICE and data has been collected for both direct emittance exchange, where the longitudinal emittance decreases, and reverse emittance exchange, where the transverse emittance decreases. The simulation studies that led to the magnet and beam configurations are presented.  
\end{abstract}
\section{Muon Ionization Cooling Experiment}
\begin{figure*}[bth]
\centering%
    \includegraphics*[width=\textwidth]{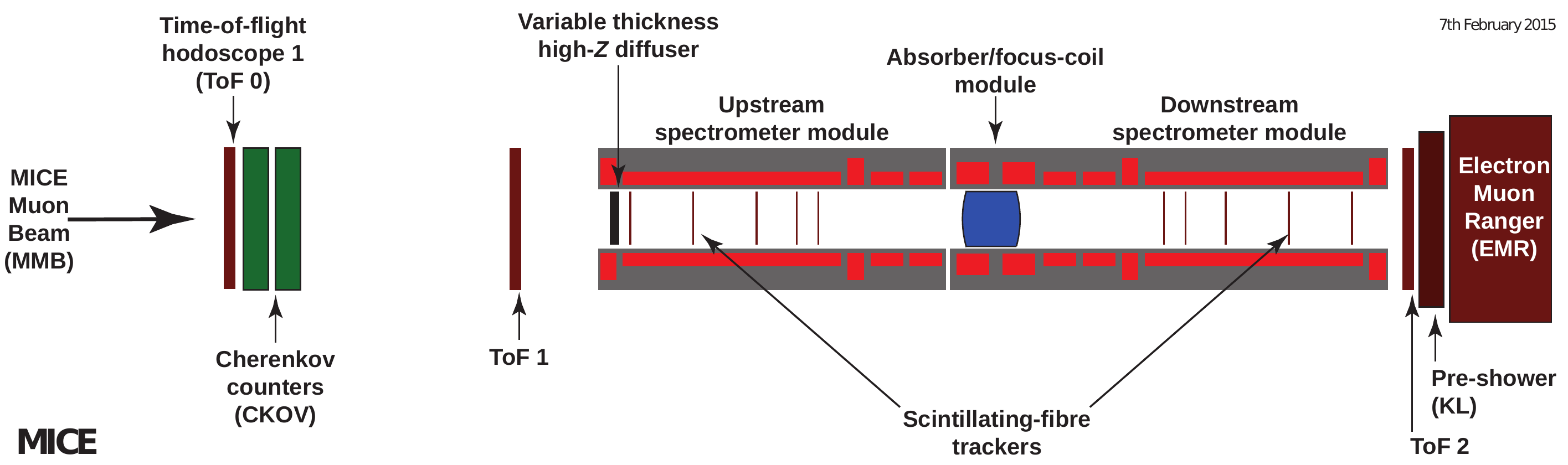}
    \caption{Schematic diagram of the Muon Ionization Cooling Experiment in its current experimental configuration.}
\label{fig:MICE}
\end{figure*}

The Muon Ionization Cooling Experiment (MICE) is the first experiment to demonstrate muon ionization cooling, the only beam cooling technique capable of reducing the muon beam phase space volume within its short lifetime~\cite{dave}. The process of ionization cooling involves the passage of muons through material where the phase-space volume or emittance of the muon beam is reduced through ionization energy loss of muons in material. MICE (Fig.~\ref{fig:MICE}) consists of two scintillating-fiber tracking detectors, one upstream and one downstream of the absorber~\cite{tracker}. Each tracker comprises five scintillating-fiber stations each with three doublet fiber layers. The muon beam cooling which results from ionization energy loss of muons in absorber is measured and compared at the locations of the tracker stations closest to the location of the absorber (referred to as tracker reference planes). The Spectrometer Solenoids housing the trackers are each made of five superconducting coils, with two used for beam matching at the absorber and three for maintaining constant solenoidal fields in the tracking volumes.
  
\section{Emittance Exchange Concept}
\begin{figure}[htbp]
\centering
    \includegraphics[width=0.6\columnwidth]{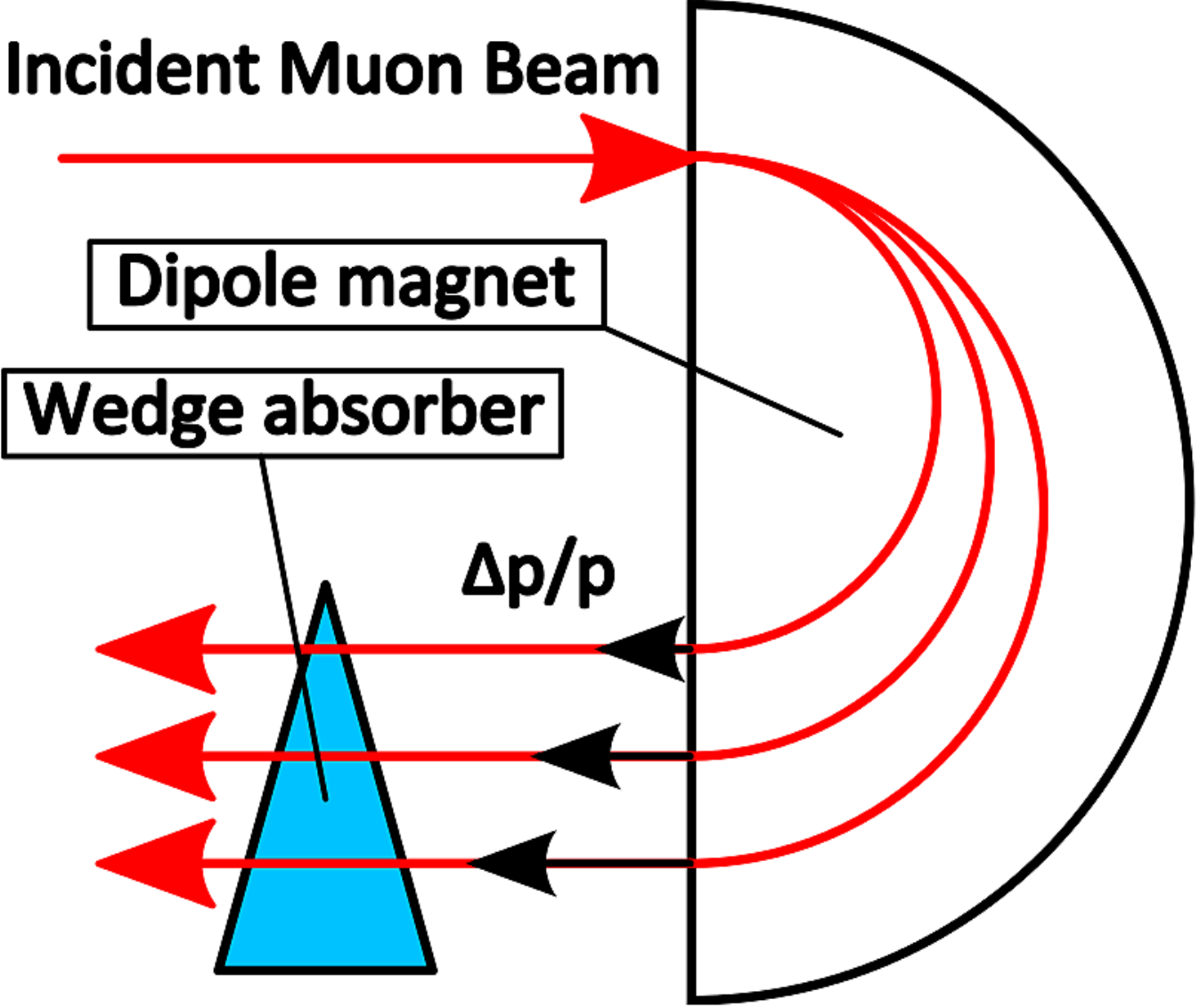}
    \caption{Diagram demonstrating the emittance exchange principle.}
\label{fig:wedge-concept}
\end{figure}

\begin{figure}[htbp]
    \centering
   \includegraphics[width=0.6\columnwidth]{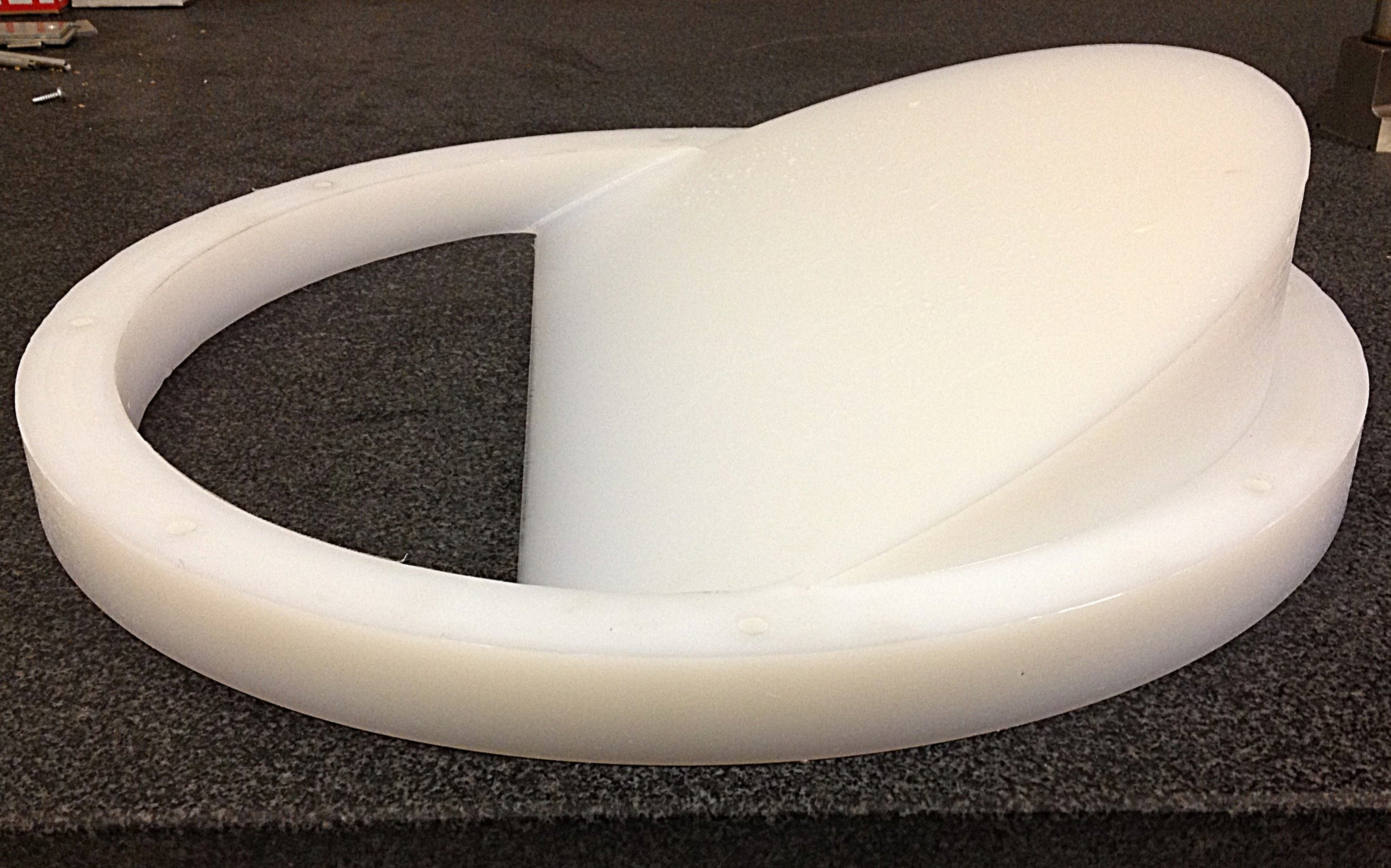}
   \caption{Half-wedge as manufactured at the University of Mississippi. The full wedge consists of two half-wedges. The resulting wedge has a 45-degree opening angle and on-axis length of 52\,mm.}
    \label{fig:half_wedge}
\end{figure}

\begin{figure}[htbp]
    \centering
   \includegraphics[width=0.6\columnwidth]{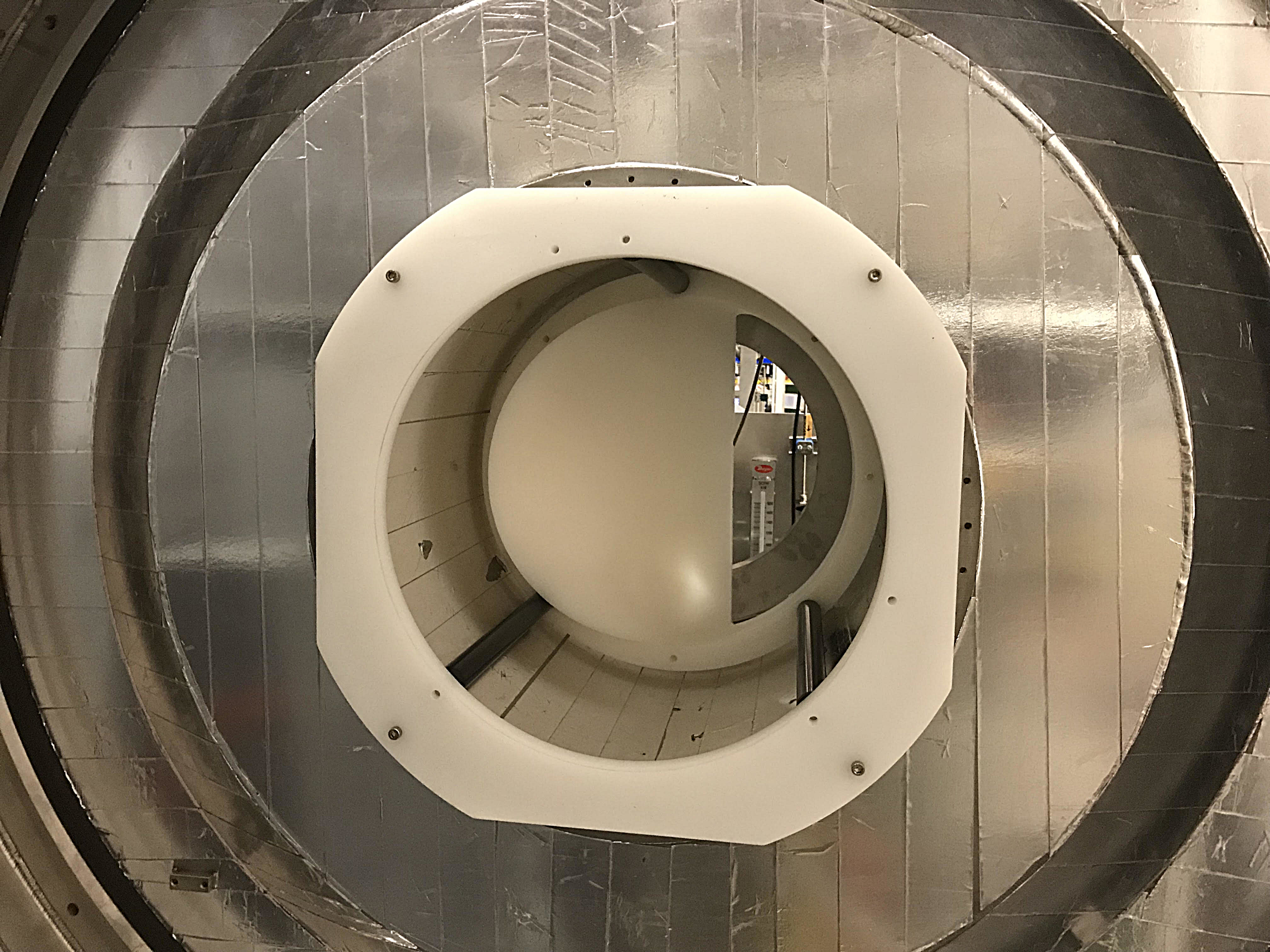}
   \caption{Wedge absorber as assembled inside the focus coil module right before data taking.}
    \label{fig:wedge}
\end{figure}

Due to the energy straggling effect in the energy loss, the use of a flat absorber in MICE can only lead to transverse phase-space volume reduction. Emittance exchange can be demonstrated in MICE by passing a dispersive beam  through a wedge absorber to cause higher energy muons  to pass through more material and lose more energy~\cite{dave,wedge_1,wedge_2,wedge_3,wedge_4,wedge_5}. This  method  causes  the  longitudinal  emittance  to  be  reduced,  enabling longitudinal cooling at the expense of transverse emittance increase (see Fig.~\ref{fig:wedge-concept}). A polyethylene wedge absorber has been fabricated (see Fig.~\ref{fig:half_wedge}) and installed in the MICE cooling channel (see Fig.~\ref{fig:wedge}). Data has been collected with wedge, and 14 million particle triggers have been recorded.
\section{Simulation Process}
The G4beamline~\cite{g4beamline} simulation package is used for tracking of muons in the MICE cooling channel as they traverse a wedge absorber. In order to preserve the desired beam dispersion in the MICE beam, an initial particle distribution is generated at the absorber and then propagated backwards to the center of the upstream tracker. The beam input emittance is $6$ mm and the reference momentum is $140$\,MeV/$c$. The beam consists of $88,000$ muons that are tracked using G4beamline across the wedge. The currents in the Spectrometer Solenoids and the absorber focus coils are one of the magnet configurations with which the wedge data was collected (with one downstream matching coil turned off). The transmission loss is $92$\% and no transmission cut is applied to discard the muons that scrape downstream of the wedge. In this simulation study, the increase in phase-space density and reduction in phase-space volume are measured as figures of merit for beam cooling. The kernel density estimation (KDE) technique, as a powerful non-parametric density estimator in muon beam cooling~\cite{KDE1,KDE2,KDE3,KDE4,KDE5}, has been applied to the simulated beam output in order to estimate the phase-space density and volume of the muon beam and identify the beam core in the four (transverse phase space), six (full transverse and longitudinal phase space), and two (longitudinal phase space) dimensional phase space. The core contour density and volume are then tracked from upstream to downstream tracker reference plane, passing through the wedge where the exchange of the longitudinal and transverse phase space is observed. 
\begin{figure}[htbp]
    \centering
   \includegraphics[width=1\columnwidth]{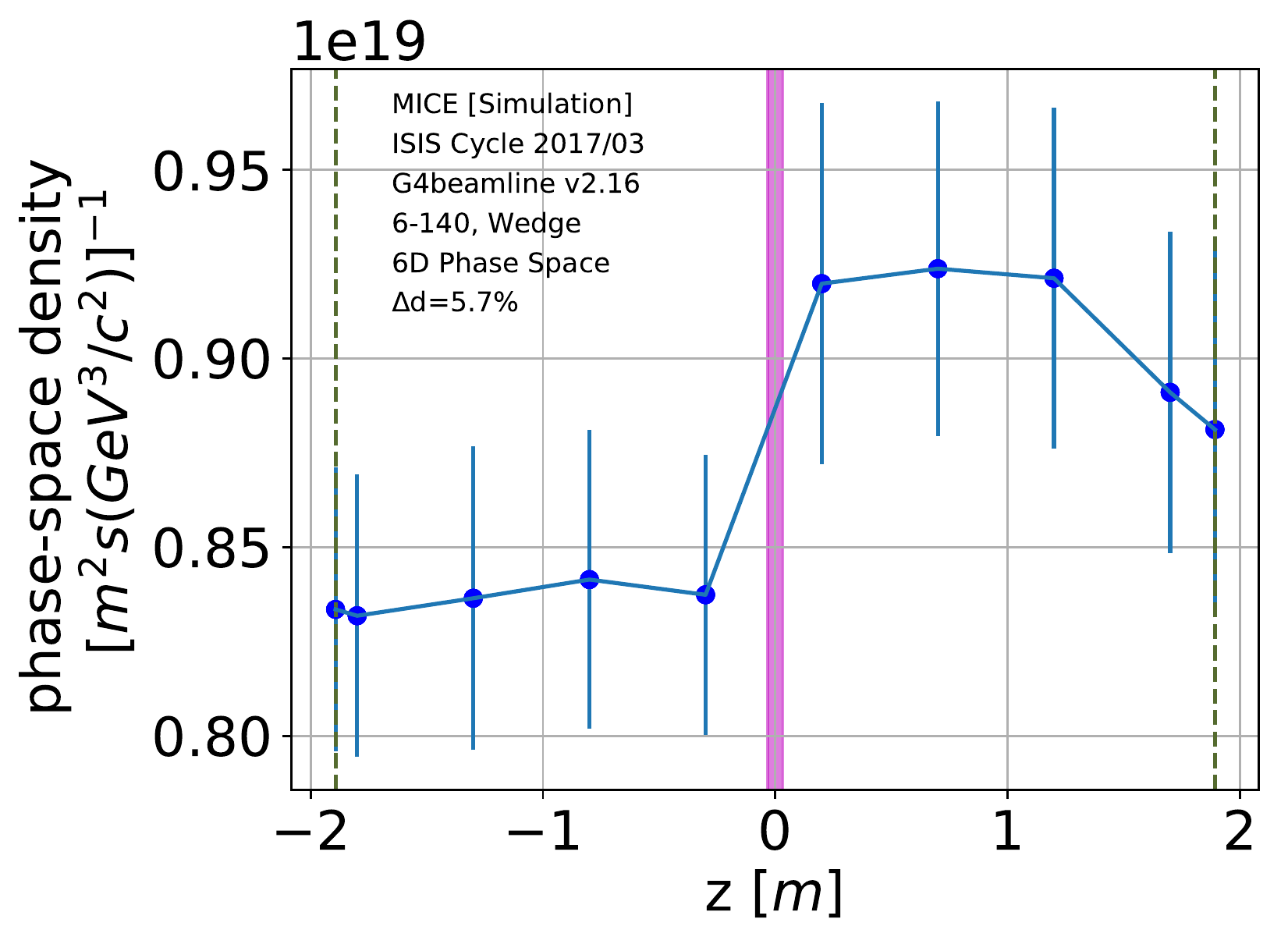}
   \caption{Six-dimensional density change of the core contour across the wedge absorber.}
    \label{fig:6d_density}
\end{figure}
\begin{figure}[htbp]
    \centering
   \includegraphics[width=1\columnwidth]{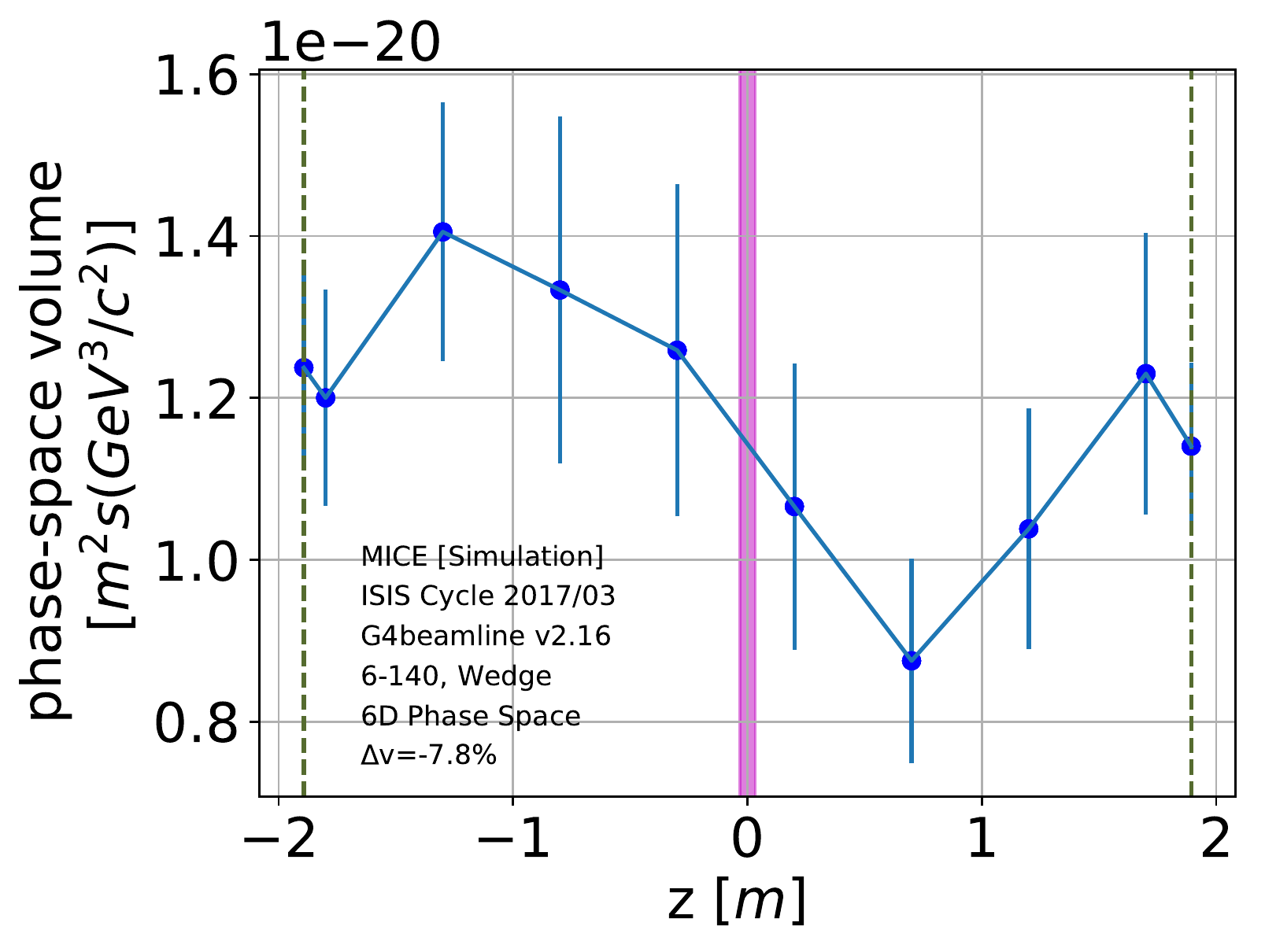}
   \caption{Six-dimensional volume change of the core contour across the wedge absorber. In 6D, the core corresponds to the $2^{\textnormal{nd}}$-percentile contour.}
    \label{fig:6d_volume}
\end{figure}
\begin{figure}[htbp]
    \centering
   \includegraphics[width=1\columnwidth]{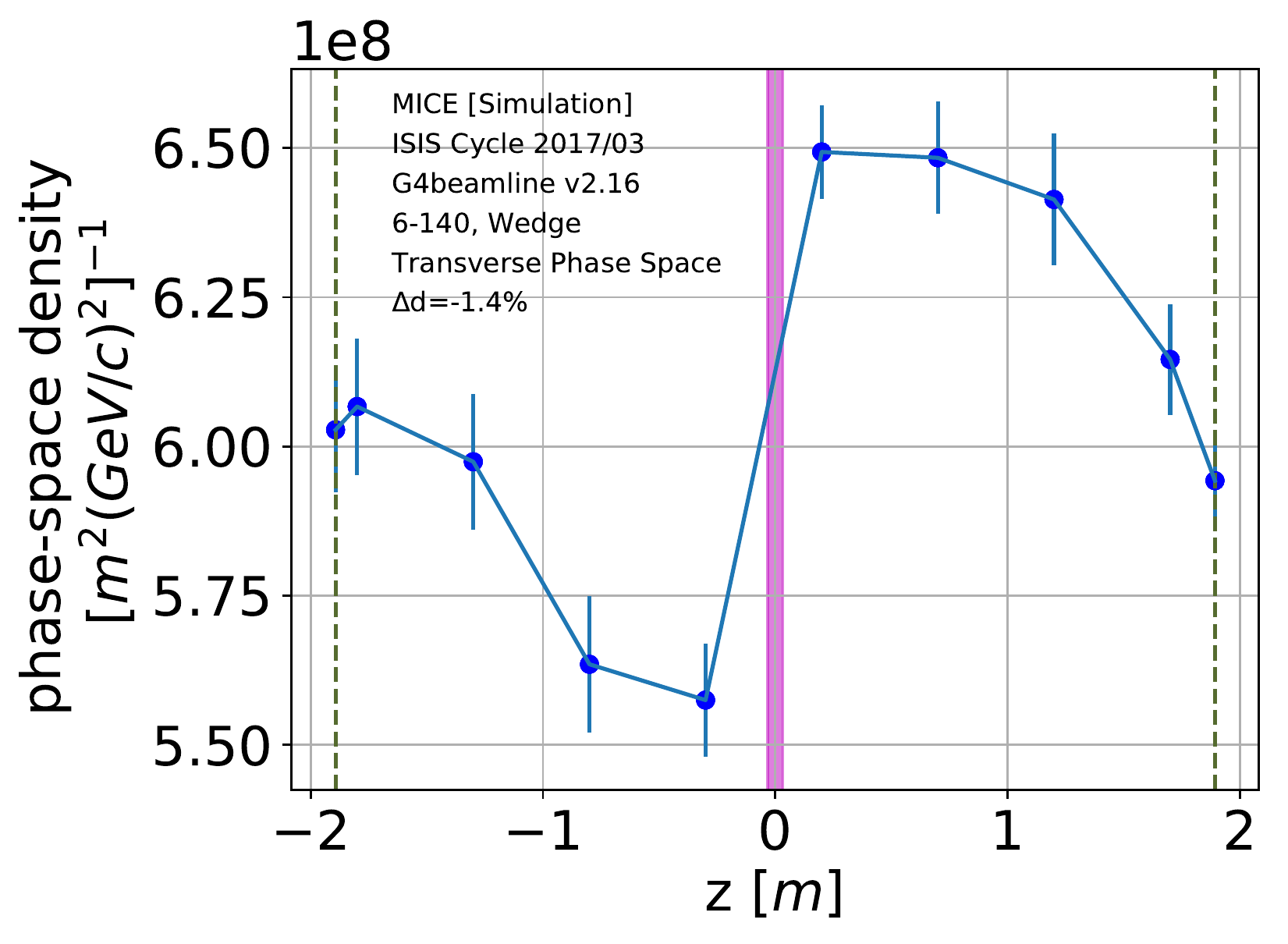}
   \caption{Core phase-space density reduction in the transverse direction is an indication of transverse heating and is an expected effect and a demonstrator of the emittance exchange process.}
    \label{fig:density}
\end{figure}
\begin{figure}[htbp]
    \centering
   \includegraphics[width=1\columnwidth]{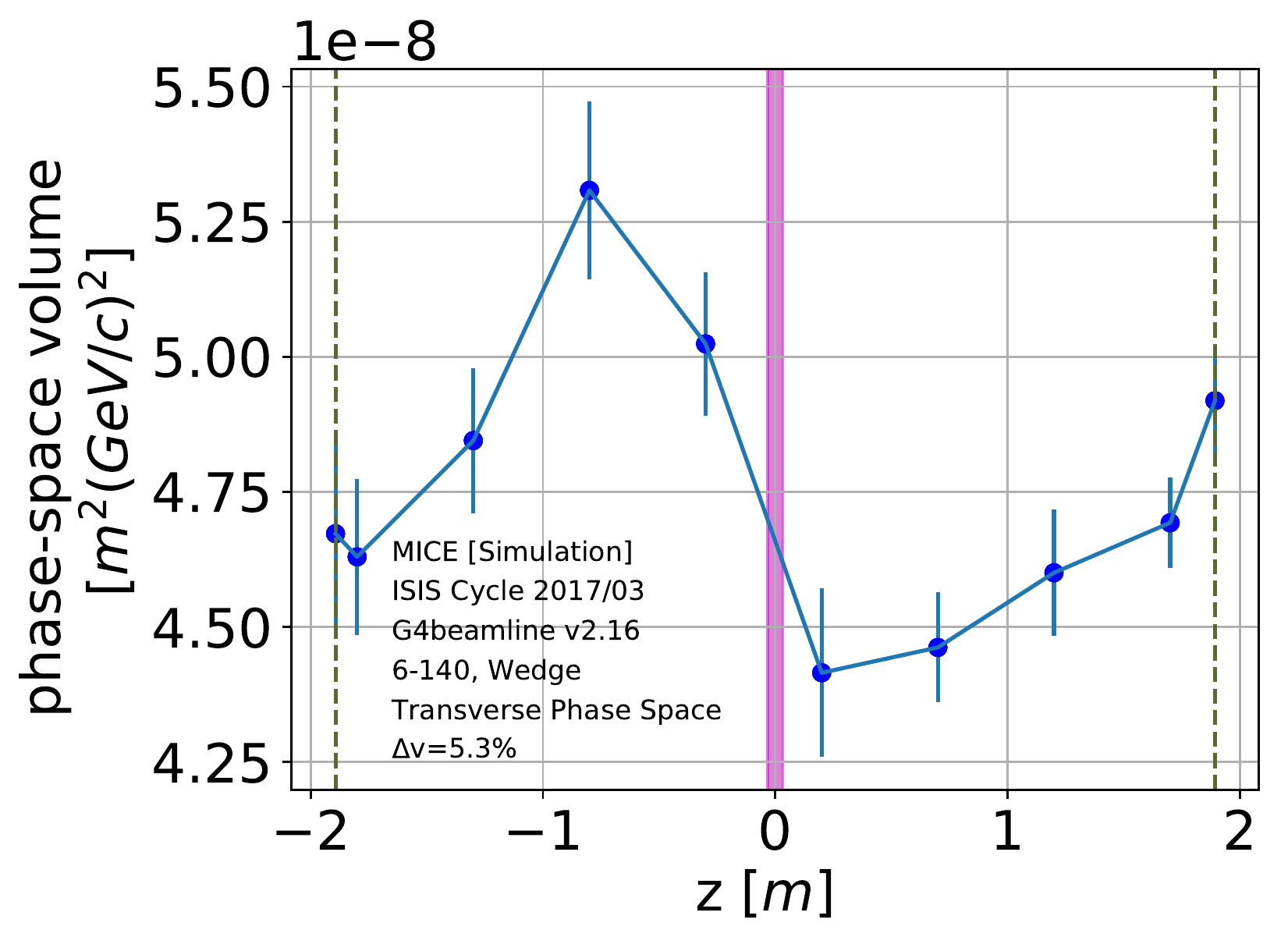}
   \caption{Phase-space volume increase in the transverse direction is an indication of the emittance exchange effect. The beam core in the four-dimensional transverse phase space corresponds to the $9^{\textnormal{th}}$-percentile contour.}
    \label{fig:volume}
\end{figure}
\begin{figure}[htbp]
    \centering
   \includegraphics[width=1\columnwidth]{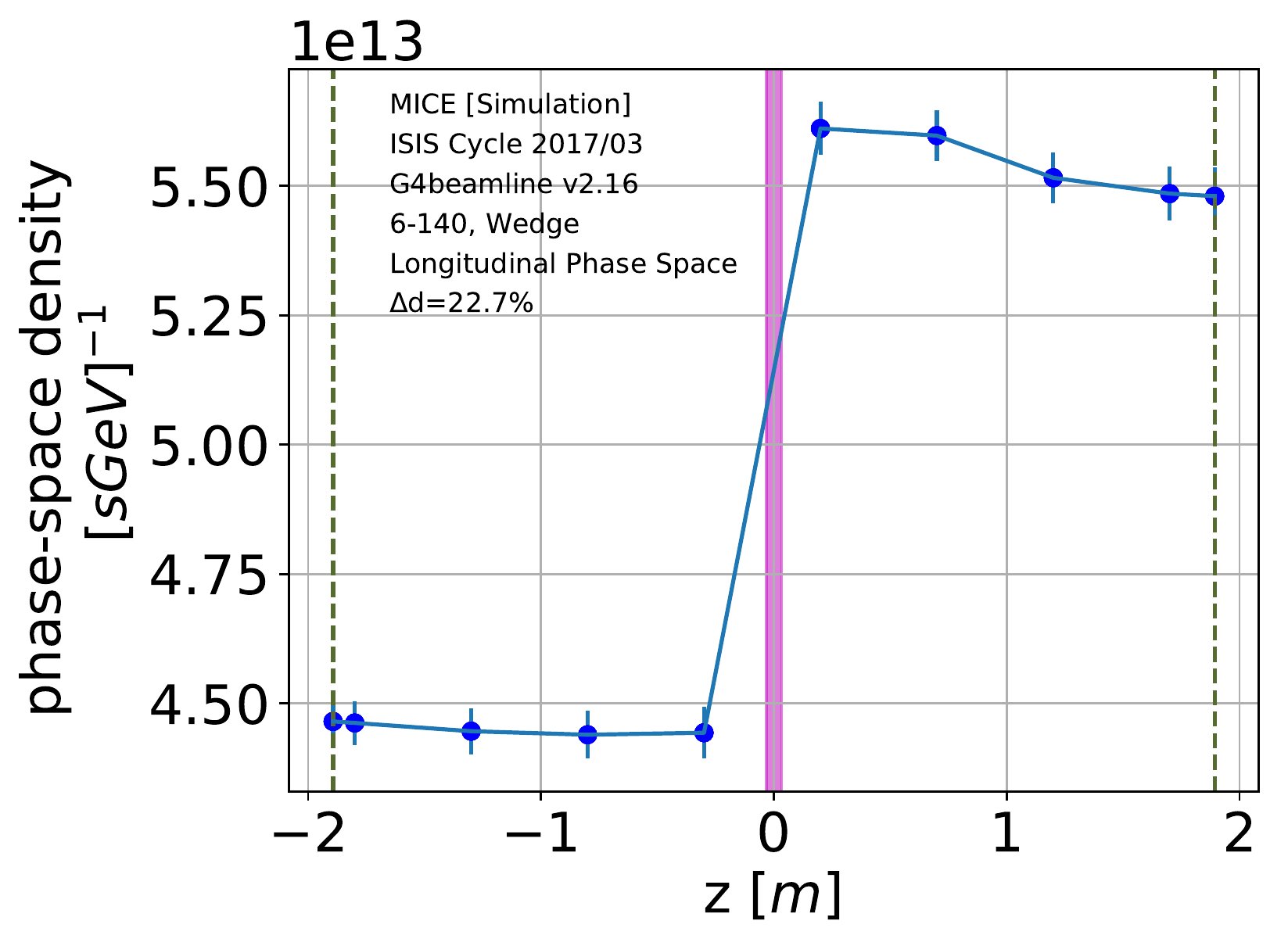}
   \caption{Increase in the longitudinal core density is an indication of cooling in the longitudinal direction.}
    \label{fig:long_density}
\end{figure}
\begin{figure}[htbp]
    \centering
   \includegraphics[width=1\columnwidth]{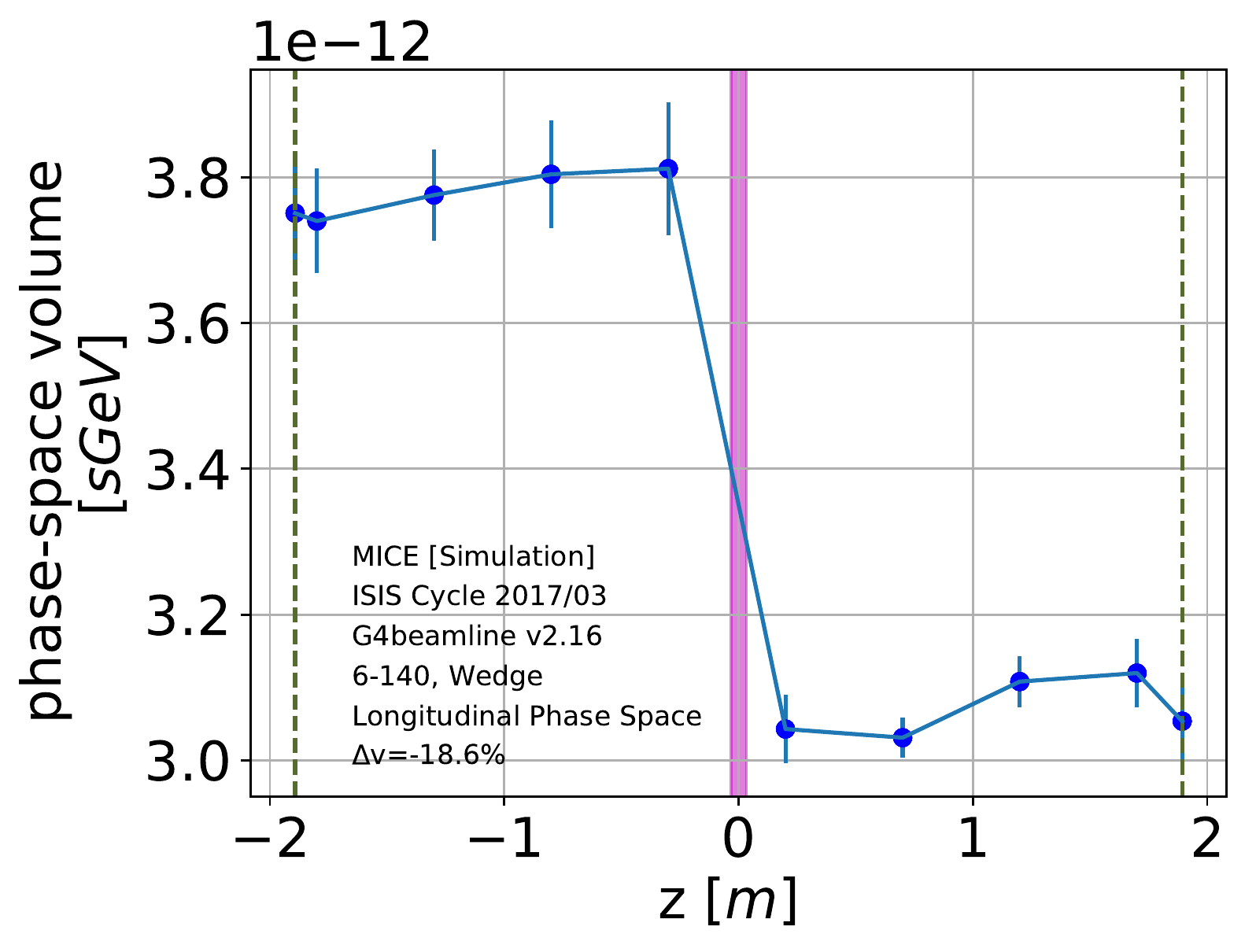}
   \caption{Reduction in the phase-space volume in the longitudinal direction is an indication of longitudinal cooling. The core contour in longitudinal direction (two dimensions) corresponds to the $24^{\textnormal{th}}$-percentile contour.}
    \label{fig:long_volume}
\end{figure}
\section{Results}
The six-dimensional phase-space evolution plots are shown in Figs.~\ref{fig:6d_density} and~\ref{fig:6d_volume}. The core of the beam in 6D corresponds to the $2^{\textnormal{nd}}$-percentile contour. Density increases and volume decreases when the beam passes through the wedge absorber. The transverse phase-space evolution plots are shown in Figs.~\ref{fig:density} and~\ref{fig:volume} where the evolutions of transverse density and volume of the core contour ($9^{\textnormal{th}}$-percentile contour in 4D) are plotted. There is a reduction in phase-space density and an increase in phase-space volume, indicating heating in the transverse direction. This heating effect combined with the longitudinal cooling seen in Figs.~\ref{fig:long_density} and~\ref{fig:long_volume} are a demonstration of the transverse-longitudinal phase-space exchange.
\section{conclusion}
The emittance exchange has been demonstrated in simulation using the kernel density estimation (KDE) technique. The application of the KDE technique to wedge data is in progress. MICE wedge data has a small natural dispersion and algorithms are being developed that can re-weight the beam distribution. 
\section{Acknowledgement}
The first author acknowledges the support that she has received from United States National Science Foundation, the Division of Physics of Beams of the American Physical Society, and TRIUMF to attend IPAC 2018.

\end{document}